\documentclass[preprint]{aastex}

\begin{document}

\title{Betelgeuse Just Isn't That Cool: Effective Temperature Alone Cannot Explain the Recent Dimming of Betelgeuse}

\author{Emily M.\ Levesque\altaffilmark{1}, Philip Massey\altaffilmark{2,3}}

\altaffiltext{1}{Department of Astronomy, University of Washington, Seattle, WA, 98195; emsque@uw.edu}
\altaffiltext{2}{Lowell Observatory, 1400 W Mars Hill Road, Flagstaff, AZ 86001; phil.massey@lowell.edu}
\altaffiltext{3}{Department of Astronomy and Planetary Science, Northern Arizona University, Flagstaff, AZ, 86011-6010}
\shortauthors{Levesque \& Massey} 
\shorttitle{Betelgeuse Isn't That Cool}

\begin{abstract}
We present optical spectrophotometry of the red supergiant Betelgeuse from 2020 February 15, during its recent unprecedented dimming episode. By comparing this spectrum to stellar atmosphere models for cool supergiants, as well as spectrophotometry of other Milky Way red supergiants, we conclude that Betelgeuse has a current effective temperature of 3600 $\pm$ 25 K. While this is slightly cooler than previous measurements taken prior to Betelgeuse's recent lightcurve evolution, this drop in effective temperature is insufficient to explain Betelgeuse's recent optical dimming. We propose that episodic mass loss and an increase in the amount of large-grain circumstellar dust along our sightline to Betelgeuse is the most likely explanation for its recent photometric evolution.
\end{abstract}

\maketitle

\section{Introduction}
As one of the closest red supergiants (RSGs) to Earth and the brightest (usually) naked-eye RSG in the night sky, Betelgeuse has long been a compelling and well-observed target for studying the late stages of massive stellar evolution. Its size ($\sim$887$\pm$203 $R_{\odot}$, Dolan et al.\ 2016) and proximity ($\sim$222$\pm$40pc, Harper et al.\ 2017) have made it an excellent target for direct-imaging observations, which have revealed variations in surface brightness typically attributed to the large convective cells expected at the surfaces of RSGs (e.g. Haubois et al.\ 2009, O'Gorman et al.\ 2017). The presence of these large-scale ``star spots" is also supported by spectropolarimetric observations of Betelgeuse's 1 G surface magnetic field (e.g. Mathias et al.\ 2018, L\'opez Ariste et al.\ 2018). Other observations of Betelgeuse have combined surface temperature and bolometric luminosity estimates to determine that Betelgeuse is the likely descendant of a main sequence star with an initial mass of $\sim$20M$_{\odot}$ (e.g. Meynet et al.\ 2013, Dolan et al.\ 2016) and the likely eventual progenitor of a Type II-P supernova. Betelgeuse's rotational velocity of $\sim$15 km s$^{-1}$, faster than the 5 km s$^{-1}$ rotation rate more typical of RSGs, has led to speculation that it may have merged with a binary companion earlier in its evolution (e.g. Wheeler et al.\ 2017). Finally, its exceptional brightness in the nighttime sky has allowed for over a century of dedicated lightcurve monitoring by amateur and professional astronomers alike.

On 2019 December 7, Guinan et al.\ (2019a) reported a $V$ = 1.12 magnitude for Betelgeuse, the faintest observed in 50+ years of continuous monitoring and considerably lower than its typical maximum brightness of V $\sim$ 0.2-0.3 mag. The decrease continued over the following two months (Guinan et al.\ 2019b, Guinan \& Wasatonic 2020a). On 2020 January 30 its $V$ magnitude was 1.614 $\pm$ 0.012, though ongoing monitoring suggested that the star's dimming could be slowing (Guinan \& Wasatonic 2020b). Betelgeuse's lightcurve shows a well-defined period of $\sim$425 $\pm$ 5 days, typically attributed to pulsations, as well as a longer-term $\sim$5.9 $\pm$ 0.5 year period; predictions suggested that if this dimming was due to a confluence of these two known periods, the star should reach minimum brightness between 2020 Feb 14-28. Publicly-released direct imaging of Betelgeuse from 2019 December using {\sc sphere}  on the Very Large Telescope showed a resolved visible-light image of Betelgeuse that included a significantly dimmer southern hemisphere as compared to 2019 January (M. Montarg\`es, private communication).

Despite considerable speculation in the popular press that Betelgeuse's visual dimming is a harbinger of an imminent supernova event, the scientific consensus (at least on social media) has settled on several less fatal explanations. One possible explanation is that variations of the large convective cells on Betelgeuse's surface could lead to a temporary decrease in the star's apparent $T_{\rm eff}$ on the timescale of weeks, shifting more blackbody emission out of the optical regime and decreasing its $V$ band. Another is that Betelgeuse could have recently undergone an episodic mass loss event, shedding mass that has condensed as circumstellar dust and is currently obscuring optical light from the star.

Here we present optical spectrophotometry of Betelgeuse taken on 2020 Feb 15 (\S2). By combining these observations with model atmospheres and existing spectra of other Milky Way RSGs, we estimate a current value for Betelgeuse's flux-weighted average surface temperature (hereafter referred to as ``$T_{\rm eff}$" for simplicity) based on the strength of the optical TiO bands (\S3). We consider these results and their implications for explaining Betelgeuse's recent dimming, and consider potential future and ongoing observations that could shed further light on Betelgeuse's current behavior and evolutionary state (\S4).

\section{Observations}
The observations were made on 2020 Feb 15 using the DeVeny low-to-moderate resolution optical spectrograph on the 4.3-meter Lowell Discovery Telescope (LDT).  As described below, we had previously observed Betelgeuse in 2004, and we strove to match the instrumental setup to that configuration.  The instrument is a conventional grating spectrometer, and was used by many observers on the Kitt Peak 0.9-m telescopes, where it was known as the White Spectrograph\footnote{When Lowell Observatory acquired the spectrograph on long-term loan, it was renamed the DeVeny Spectrograph in honor of James DeVeny, the late head of their instrument support group.} As part of our upgrade, a new CCD camera was installed, using an e2v CCD42-10 deep depletion device with a 4-layer AR coating. The CCD has 2048$\times$512 13.5~$\mu$m pixels.  

For the observations of Betelgeuse, a 500 line mm$^{-1}$ grating (formerly known as KPNO 240) was used.  The slit was opened to 3\farcs0 to allow good spectrophotometry, and all observations were made with the slit aligned at the parallactic angle.  The CCD was binned to 5 (spatial axis) $\times$2 (wavelength axis), resulting in a scale of 1\farcs7 pixel$^{-1}$ and a dispersion of 2.67\AA\ pixel$^{-1}$.  The wavelength coverage was 4000-6700\AA\ with a resolution of 8.0\AA.  Wavelength calibration was by means of Cd, Ar, and Hg comparison spectrum.  Pixel-to-pixel variations were removed by flat-fielding by observations of a calibration screen.  Spectral reductions were performed using {\sc iraf}. 

Despite its recent dimming, Betelgeuse is still one of the brightest stars in the sky, and so observations had to be performed with a 7.5 mag neutral density (ND) filter.  The ND filter is located above the slit, thus easing issues of acquisition with the CCD viewing camera.  Such filters are in fact never ``neutral", and a comparison of flat-field exposures with and without filtering shows a drop in transmission of $2\times$ from the red to the blue side of the spectrum.  Accordingly, we observed (bright) spectrophotometric standards through the same ND filter, specifically HR 718, HR 1544, and HR 3454, all chosen from Hamuy et al. \ (1992, 1994).  The airmasses of the spectrophotometric observations were 1.1-1.4, while that of Betelgeuse was 1.2.  We assumed the standard {\sc iraf} Kitt Peak extinction coefficients, but since the standards were observed at comparable air masses as the science target, any deviations from those values is immaterial.  The RMS residuals from the standard star fits were 0.02~mag after a grey shift.  We expect relative fluxes to be good to a few percent based upon the agreement of the three standards.

\section{The Effective Temperature of Betelgeuse in 2020}
In Levesque et al.\ (2005, hereafter Paper I) we measured $T_{\rm eff}$ for 74 Galactic RSGs, including Betelgeuse, by comparing the strengths of the stars' optical TiO bands to the predictions of the latest generation of MARCS stellar atmosphere models (e.g. Gustafsson et al.\ 2008). The strength of the TiO bands has long been established as a means of estimating the $T_{\rm eff}$ of RSGs to high precision. Although some have questioned the use of TiO band strengths as a means of accurately measuring the $T_{\rm eff}$ for RSGs (e.g., Davies et al. 2013), the spectral subtypes themselves are defined based upon the strengths of these bands. Early work by Wing (1992) established a narrow-band filter system that estimated $T_{\rm eff}$ for cool stars based on the strength of the 7054\AA\ TiO band. Estimates of $T_{\rm eff}$ for RSG based on the strengths of the TiO bands show excellent agreement with the temperatures of RSGs predicted by single non-rotating, single rotating, and binary evolutionary models as well as the metallicity-dependent shift of the Hayashi limit seen in both theoretical models and observed stellar populations (e.g. Elias et al.\ 1985, Massey \& Olsen 2003, Paper I, Levesque et al.\ 2006, Massey et al.\ 2009, Levesque \& Massey 2012, Drout et al.\ 2012, Ekstr\"om et al.\ 2012, Georgy et al.\ 2013, Dorda et al.\ 2016, Eldridge \& Stanway 2016, Levesque 2017, Levesque 2018). The MARCS stellar atmosphere models and PHOENIX stellar atmosphere models (Lancon et al.\ 2007) both show good agreement with observed TiO bands strengths and their evolution as a function of temperature, and 3D stellar atmosphere models for RSGs also note that the TiO bands are temperature-sensitive (e.g. Chiavassa et al.\ 2011). While other methods also provide good estimates of RSG $T_{\rm eff}$ (most notably color-based methods that combine optical and near-IR colors, such as the $(V-K)_0$ color index; see Massey et al.\ 2009, Levesque 2018), the optical TiO bands currently remain the best means of measuring $T_{\rm eff}$ for a RSG\footnote{That said, it is worth reminding
the reader that our method relies upon measuring the integrated spectrum over a disk that doubtless contains regions of different temperatures to a 2D model atmosphere, where all points on the disk would be of uniform value.  Thus our effective temperature is basically a flux-weighted mean.}.

We begin by comparing our 2020 Betelgeuse spectrum to our previous observation taken on 2004 March 7-8. At that time Betelgeuse had a $V$ mag of $\sim$0.5, $\sim$1.1 mag brighter than its brightness on 2020 Feb 15, and a $T_{\rm eff}$ = 3650 $\pm$ 25 K. A comparison of our two SEDs for Betelgeuse is shown in Figure 1. The 2020 flux is notably lower than the 2004 flux; however, the overall shape of the two SEDs is very similar, with a slightly larger excess in the blue evident in the 2020 spectrum. Figure 2 directly compares the depths of the TiO absorption bands in the normalized 2004 and 2020 spectra. The 2020 spectrum appears to have slightly stronger TiO bands, and thus a slightly cooler $T_{\rm eff}$. The wavelength range of our 2020 spectrum covers the 4761\AA, 4954\AA, 5167\AA, 5448\AA, 5847\AA, and 6158\AA\ bands; all bands are slightly wider and deeper than the 2004 spectrum. As detailed in Levesque (2017), these bands are all $T_{\rm eff}$-sensitive, and the bluer TiO bands in particular are sensitive to changes in temperature below $\sim$3700 K.

We can also compare our 2020 Betelgeuse spectrum to several other typical Galactic RSGs with well-determined temperatures from Paper I. Figure 3 shows our 2020 Betelgeuse spectrum as compared to three other Galactic RSGs: HD 94096 ($T_{\rm eff}$ = 3650 K), BD+60$^{\circ}$ 2634 ($T_{\rm eff}$ = 3600 K), and HD 14488 ($T_{\rm eff}$ = 3550 K); all $T_{\rm eff}$ values have fitting uncertainties of $\pm$25 K. This comparison shows that the TiO bands in our 2020 Betelgeuse spectrum are weaker than those seen in HD 14488, but stronger than those seen in HD 94096; they agree best with the spectrum of BD+60$^{\circ}$ 2634. This comparison with other Galactic RSGs also supports a slightly cooler $T_{\rm eff}$ for Betelgeuse in 2020 as compared to 2004.

Finally, we can fit our 2020 Betelgeuse spectrum using the MARCS stellar atmosphere models. Figure 4 shows the best-fit MARCS stellar atmosphere models to both our 2004 and 2020 spectra: the 2004 spectrum has a best-fit model of 3650 K, while the 2020 spectrum has a best-fit model of 3625 K. However, it is worth noting that while the MARCS models do an excellent job of fitting most TiO bands they tend to slightly underestimate the peak flux of the 5167\AA\ bandhead across all RSG atmosphere models as $T_{\rm eff}$ decreases. The error bars and 5167\AA\ bandhead underestimate suggest that the MARCS model fit to our 2020 Betelgeuse spectrum may risk slightly overestimating its temperature. By combining our MARCS model fit and the direct comparisons with Galactic RSGs, we therefore estimate that our 2020 spectrum of Betelgeuse has a $T_{\rm eff}$ = 3600 $\pm$ 25 K.

It is worth noting that this is in generally good agreement with the $T_{\rm eff}$ estimates of 3580 K on 2019 Dec 7 and 3565 K on 2020 Jan 31 published by Guinan et al.\ (2019a) and Guinan \& Wasatonic (2020b), respectively, based on narrow-band Wing TiO photometry (it is also interesting to note that these two $T_{\rm eff}$ estimates are extremely close despite a decrease of 0.52 mag in Betelgeuse's $V$ magnitude between the two measurements). Guinan et al.\ (2019a) also quotes a ``normal" $T_{\rm eff}$ of 3660 K based on Wing TiO photometry, in good agreement with our determination of Betelgeuse's $T_{\rm eff}$ from 2004.

Our best-fit MARCS stellar atmosphere models from 2004 and 2020 also illustrate a surprisingly consistent $A_V$ along our line of sight to Betelgeuse; in both 2004 and 2020 the best-fit models adopt a surface gravity of log $g$ = 0.0 and an $A_V = 0.62$, assuming a standard Cardelli et al.\ (1989) reddening law. However, our observed 2020 spectrum also shows a slight flux excess blueward of 4500\AA\ when compared to the best-fit MARCS model, a mismatch that is notably less pronounced when examining the best-fit model of our 2004 spectrum. Massey et al.\ (2005) found that significant excesses in the near-UV spectra of RSGs were likely due to scattering of the star's light by circumstellar dust with a larger average dust grain size.

In summary, there has been no apparent change in the Cardelli-law dust content along our line of sight to Betelgeuse, and the star's $T_{\rm eff}$ has dropped by only $\sim$50 K when comparing the 2004 spectrum (when $V \sim$ 0.5) to our new 2020 spectrum (when $V \sim$ 1.6). When considering the $\pm$25 K precision of our TiO band fitting method, Betelgeuse's $T_{\rm eff}$ during its current optical dimming episode has decreased by at most 100 K; at minimum it has not changed at all.

\section{Implications and Future Work}
It seems clear from our recent optical spectrophotometry that Betelgeuse's $T_{\rm eff}$ has not decreased significantly in connection with its recent visible decrease in brightness. If Betelgeuse's lower $V$ band magnitude was the result of a lower $T_{\rm eff}$ - caused, for example, by a temporarily low apparent $T_{\rm eff}$ due to surface convection effects - we would expect to see substantially strong TiO bands and a much lower $T_{\rm eff}$ from our observations. These results suggest that a temporary ``cold" period on the surface of Betelgeuse due to convective turnover is likely {\it not} the primary cause of Betelgeuse's recent dimming. Still, it is worth considering whether even this relatively small change in $T_{\rm eff}$ is sufficient to explain Betelgeuse's recent variability in $V$. A lower $T_{\rm eff}$ corresponds to a longer peak wavelength, shifting more of the star's light out of the $V$ band; it's certainly possible to explain a decrease in $V$ magnitude with a drop in a star's $T_{\rm eff}$.

Following the $V$ band bolometric corrections implied by the MARCS stellar atmosphere models (see Paper I, Levesque et al.\ 2006), where: $$BC_V = -298.854+217.532(T_{\rm eff}/1000 K)-53.14(T_{\rm eff}/1000 K)^2+4.34602(T_{\rm eff}/1000 K)^3$$ a drop from 3650 K to 3600 K in $T_{\rm eff}$ would correspond to a decrease in $V$ of only $\sim$0.17 mag. Even adopting the maximum possible decrease from our results, from 3650 K to 3550 K, corresponds to a decrease of $V$ of $\sim$0.38 mag. Given that Betelgeuse has shown a decrease of $V \sim 1.1$ mag between our 2004 and 2020 observations, our observed $T_{\rm eff}$ cannot explain Betelgeuse's lightcurve over the past few months.

This suggests that the recent dimming of Betelgeuse must be due, at least in part, to some substantial effect other than a change in $T_{\rm eff}$. Our best-fit models suggest at first glance that the reddening around Betelgeuse appears unchanged; however, this method assumes line-of-sight dust that obeys the Cardelli et al.\ (1989) reddening law. In truth, observations have demonstrated that the dust produced by RSG mass loss has a much larger grain size. Snow et al.\ (1987) found evidence of large siliceous circumstellar dust grains around the RSG binary $\alpha$ Sco. Scicluna et al.\ (2015) estimated a dust grain size of 0.5$\mu$m around the dust-enshrouded RSG VY CMa, and Massey et al.\ (2006) identified the UV excess characteristic of large-grain circumstellar dust in optical spectrophotometry of VY CMa. Haubois et al.\ (2019) recently estimated grain sizes of $\sim$0.3$\mu$m in a dust halo located only half a stellar radius above the photosphere of Betelgeuse, the region where most extinction from circumstellar dust will occur (see also Massey et al.\ 2005). Such large-grain dust has been known to form from some classical novae as well (see, e.g., Shore et al. 1994). Larger grain sizes correspond to extinction that is ``gray", absorbing light across the optical spectrum rather than preferentially absorbing bluer wavelengths.

The dimming of Betelgeuse on a timescale of months also agrees well with existing research on mass loss and dust production in RSGs. Observations of circumstellar dust shells around RSGs show evidence of episodic and asymmetric mass loss (e.g. Danchi et al.\ 2004, Smith et al.\ 2001, Schuster et al.\ 2006, Ohnaka et al.\ 2008, Scicluna et al.\ 2015). Kervella et al.\ (2018) observed evidence of asymmetric mass loss and dust formation from Betelgeuse (see also Kervella et al.\ 2016). Levesque et al.\ (2007) and Massey et al.\ (2007) identified several RSGs in the Magellanic Cloud whose circumstellar reddening appeared to change on the timescale of months (though it's worth noting that those stars, unlike Betelgeuse, also exhibited significant variations in $T_{\rm eff}$).

We propose that, based on our spectrophotometry, an increase in large-grain gray dust due to recent episodic mass loss from Betelgeuse is the best explanation for its recent lightcurve evolution. This explanation agrees with the lack of significant changes seen in the star's $T_{\rm eff}$ as well as with the surprising consistent amount of Cardelli-law line of sight dust towards Betelgeuse combined with an increase in flux at the bluest wavelengths in our 2020 spectrum. The 2019 January VLT {\sc sphere} resolved image, which clearly shows extensive ``dark" regions across the southern portion of the star's disk that we now know are not accompanied by a notable change in $T_{\rm eff}$, is also consistent with our proposed explanation of gray extinction.

However, more observations in the coming weeks and months are required to test this hypothesis. UV observations could shed light on the outer layers of Betelgeuse as well as any unusual signatures of the dust reflection nebula effects discussed in Massey et al.\ (2005). Direct imaging and polarimetry of Betelgeuse's surface and circumstellar environment could illuminate the geometry of the star's circumstellar environment. Mid-infrared observations could potentially be used to quantify the mass, distribution, and composition of circumstellar dust (e.g. Verhoelst et al.\ 2009). Finally, ongoing photometric observations in both the optical and near-IR regimes are crucial for continued monitoring of Betelgeuse's behavior and evolution. Combined, these observations will be able to test the circumstellar-dust explanation for Betelgeuse's unprecedented dimming and place the behavior of this nearby star in context with other RSGs, allowing it to serve as a valuable example of post-main-sequence (and pre-supernova) massive stellar evolution.

\acknowledgements We thank the anonymous referee for constructive and helpful comments that have improved the quality of this manuscript. These results made use of the Lowell Discovery Telescope at Lowell Observatory. Lowell is a private, non-profit institution dedicated to astrophysical research and public appreciation of astronomy and operates the LDT in partnership with Boston University, the University of Maryland, the University of Toledo, Northern Arizona University and Yale University. The upgrade of the DeVeny optical spectrograph has been funded by a generous grant from John and Ginger Giovale and by a grant from the Mt. Cuba Astronomical Foundation. EML is supported by a Cottrell Scholar Award from the Research Corporation for Scientific Advancement. PM acknowledges support from the National Science Foundation (AST-1612874).

\begin{figure}
\plotone{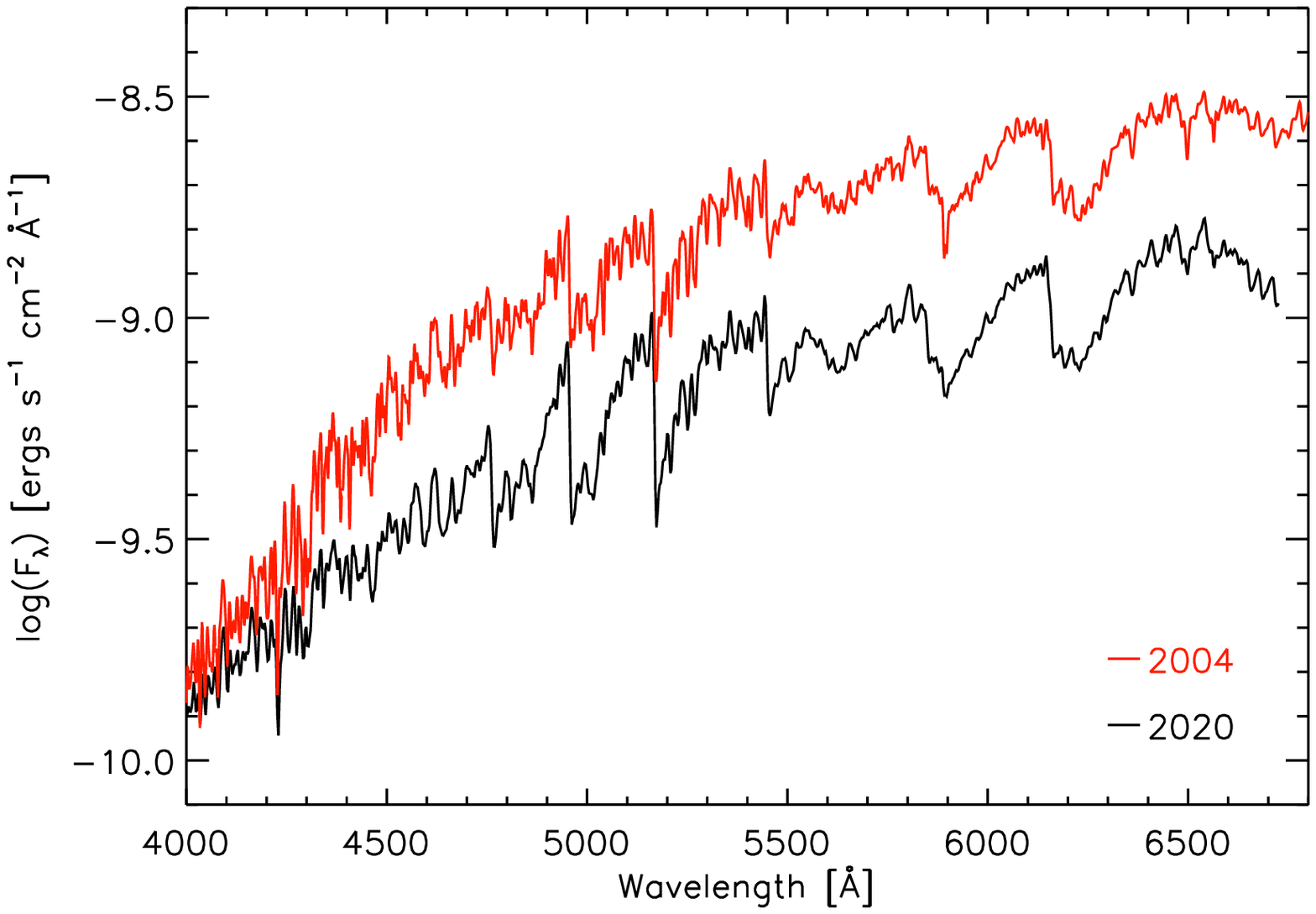}
\caption{Optical spectrophotometry of Betelgeuse from 2020 (black) and 2004 (red).}
\end{figure}

\begin{figure}
\plotone{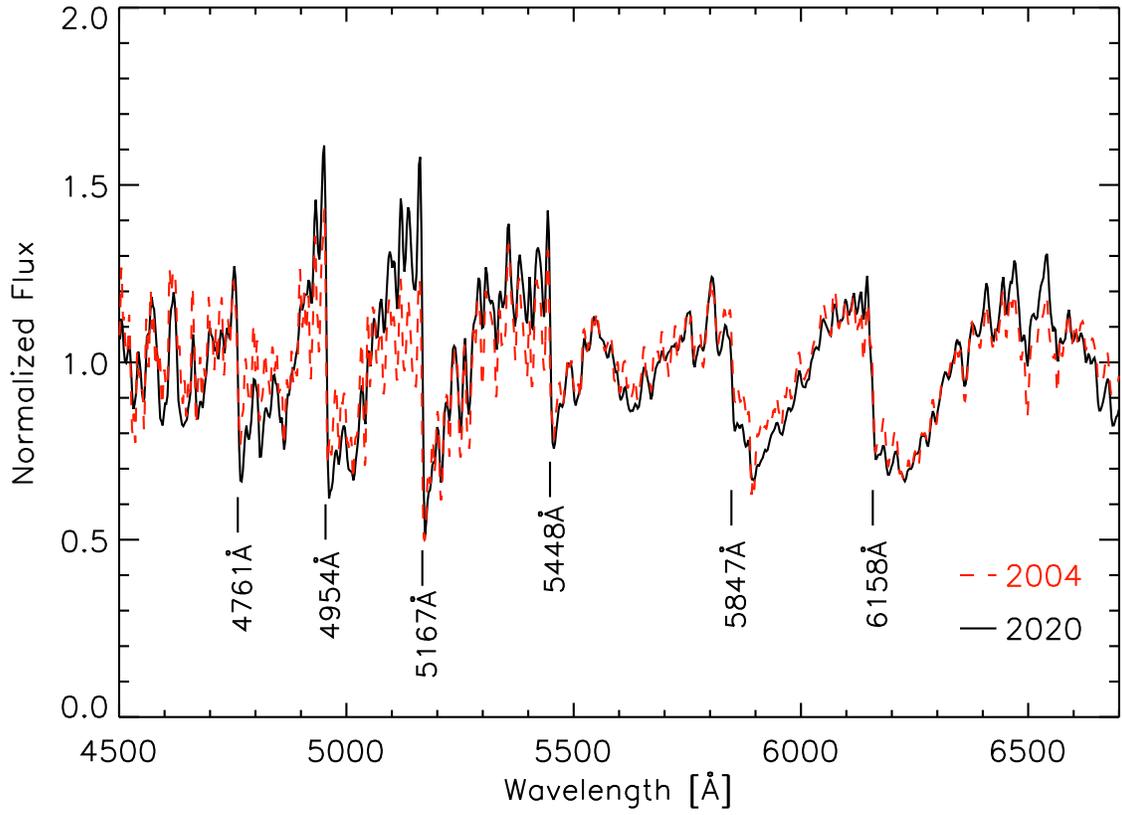}
\caption{Normalized and continuum-flattened spectra of Betelgeuse from 2020 (black) and 2004 (red dashed), illustrating the relative depths of the $T_{\rm eff}$-sensitive TiO bands (labeled) in both spectra.}
\end{figure}

\begin{figure}
\epsscale{0.49}
\plotone{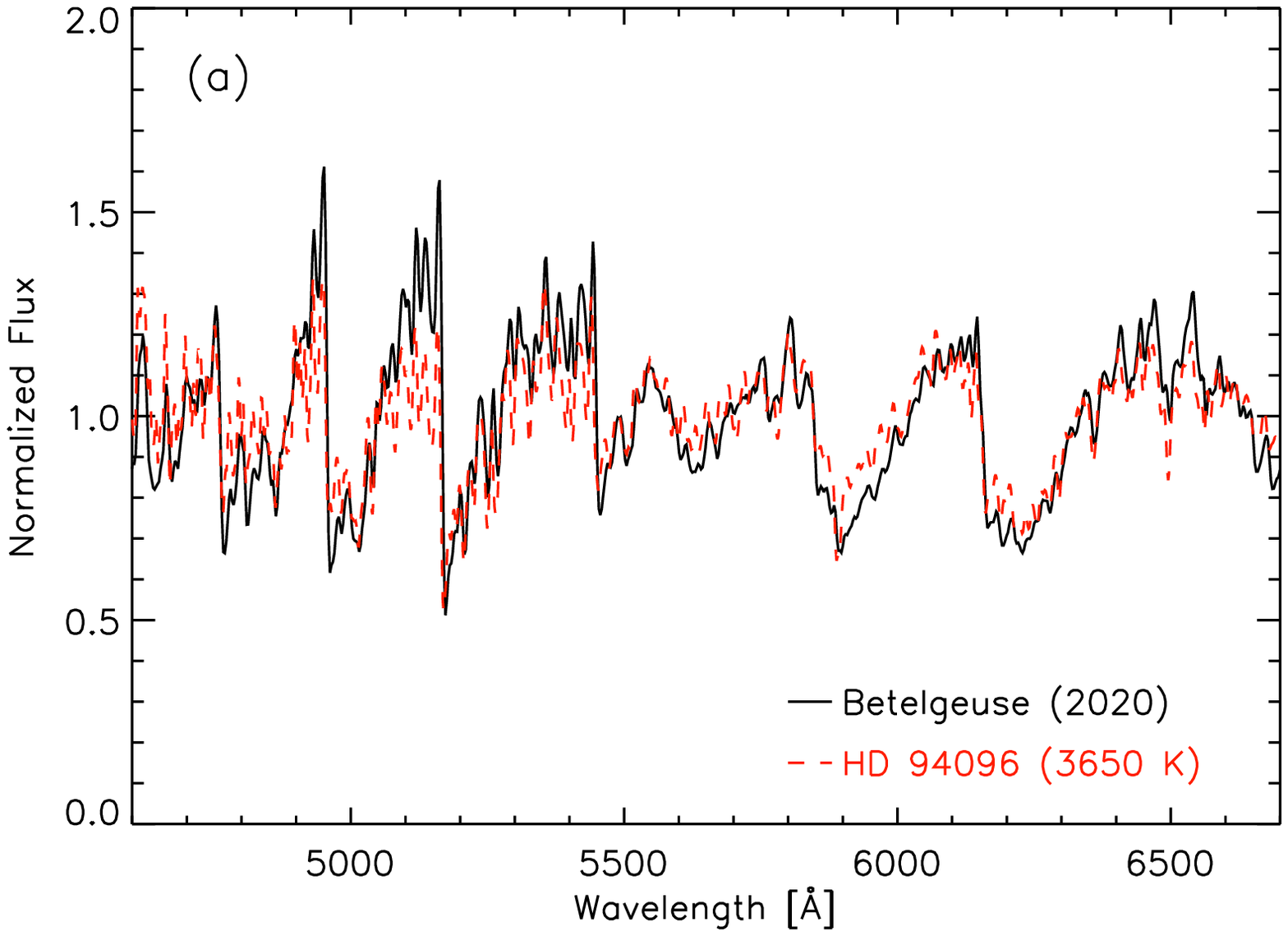}
\plotone{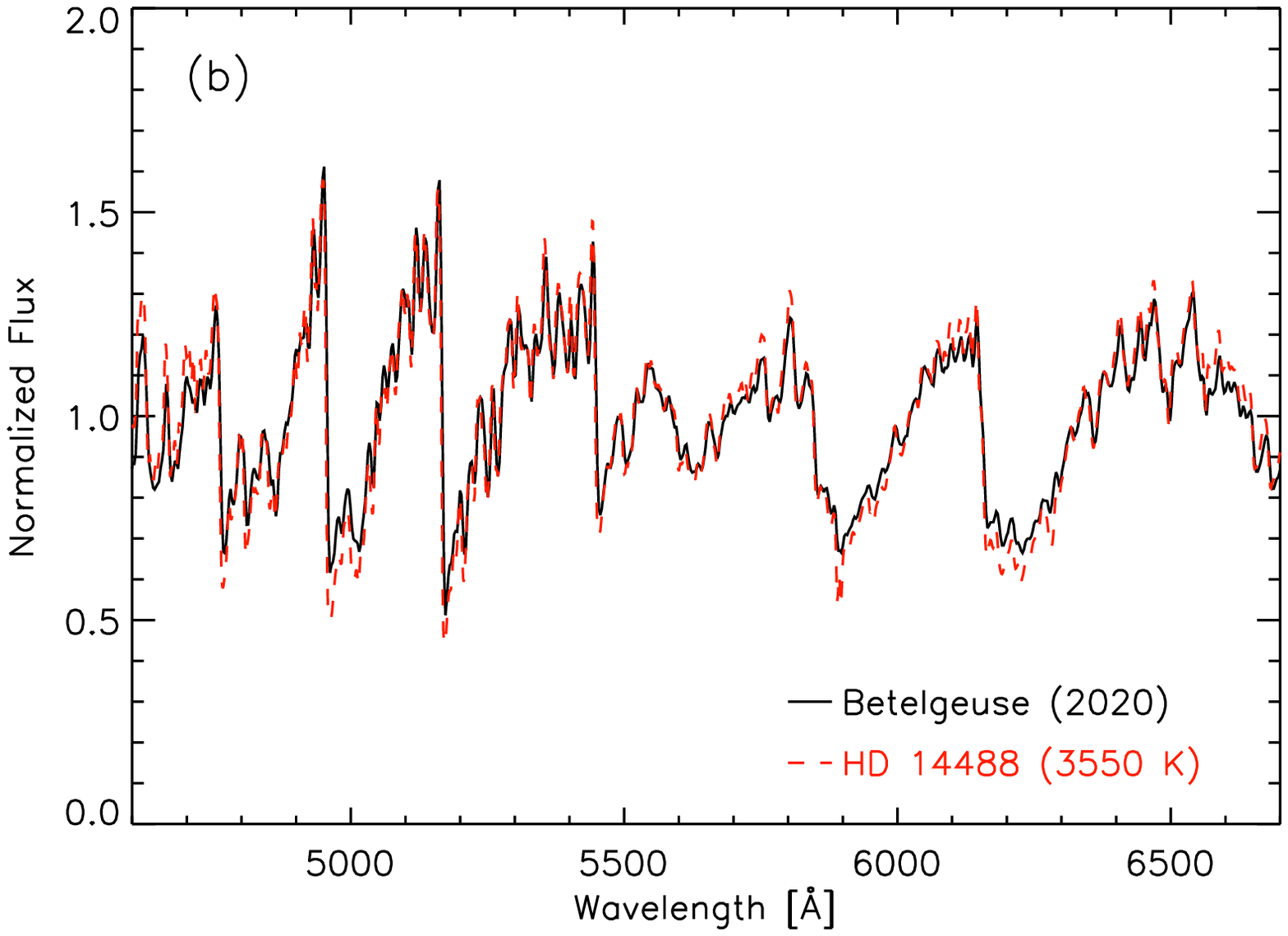}
\plotone{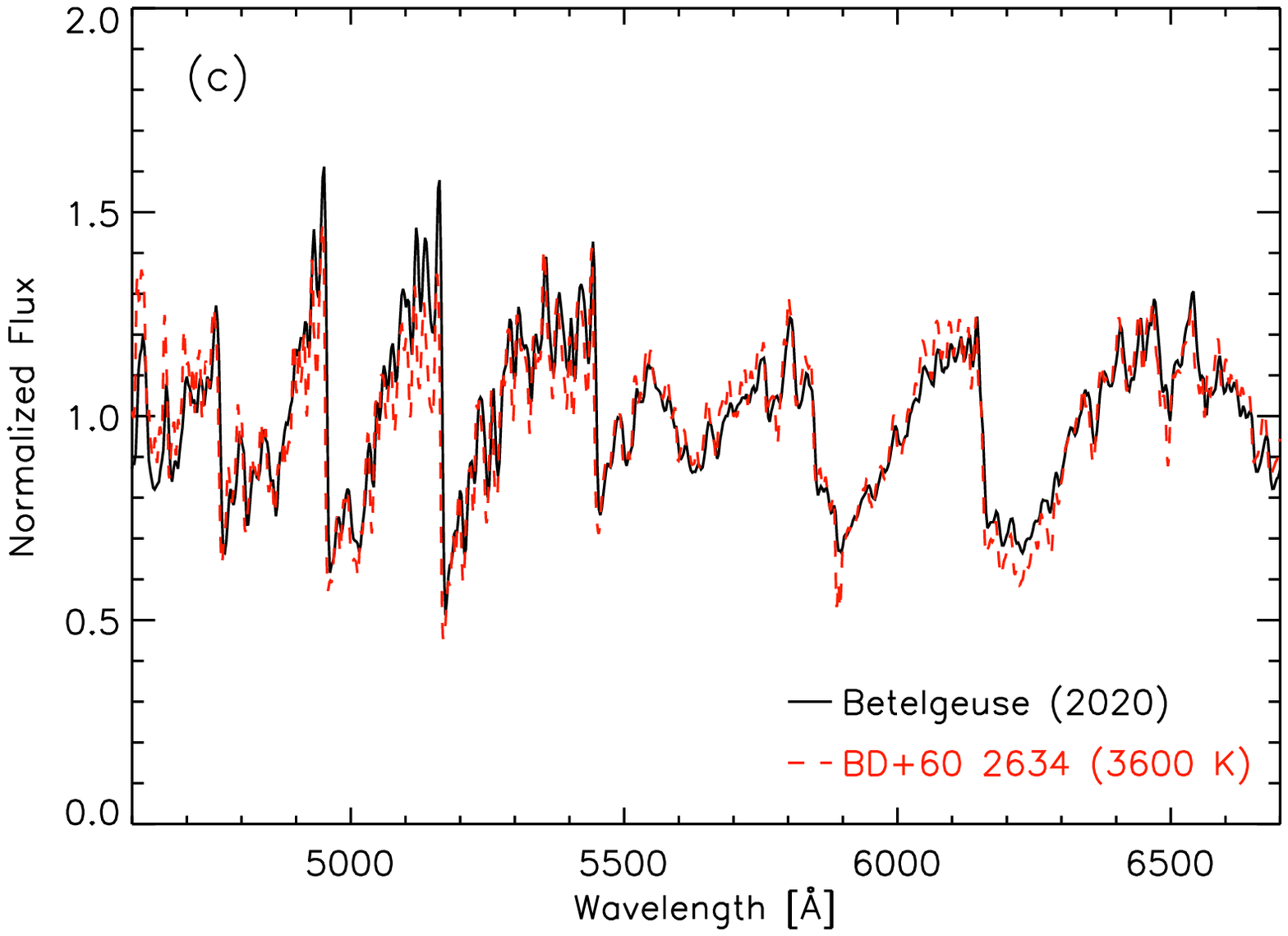}
\plotone{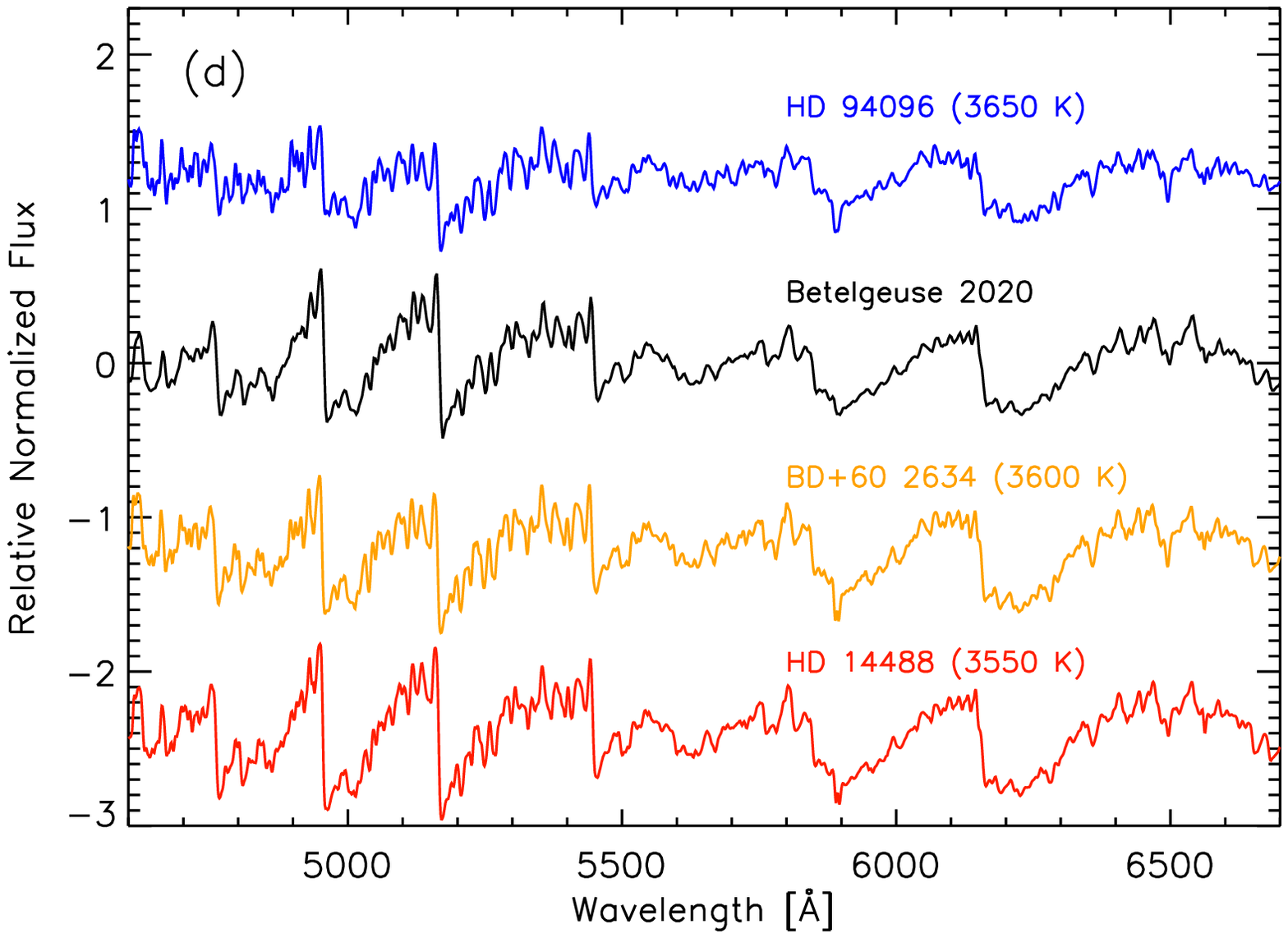}
\caption{Comparison of the TiO band depths in our normalized 2020 Betelgeuse spectrum (black) to the Galactic RSGs HD 94096 (a), HD 14488 (b), and BD+60$^{\circ}$ 2634 (c). In figure (d) all four spectra are stacked and compared. The TiO bands in our 2020 Betelgeuse are stronger that those seen in HD 94096 ($T_{\rm eff}$ = 3650 K), weaker than those seen in HD 14488 ($T_{\rm eff}$ = 3550 K), and comparable to those seen in BD+60$^{\circ}$ 2634 ($T_{\rm eff}$ = 3600 K).}
\end{figure}

\begin{figure}
\plotone{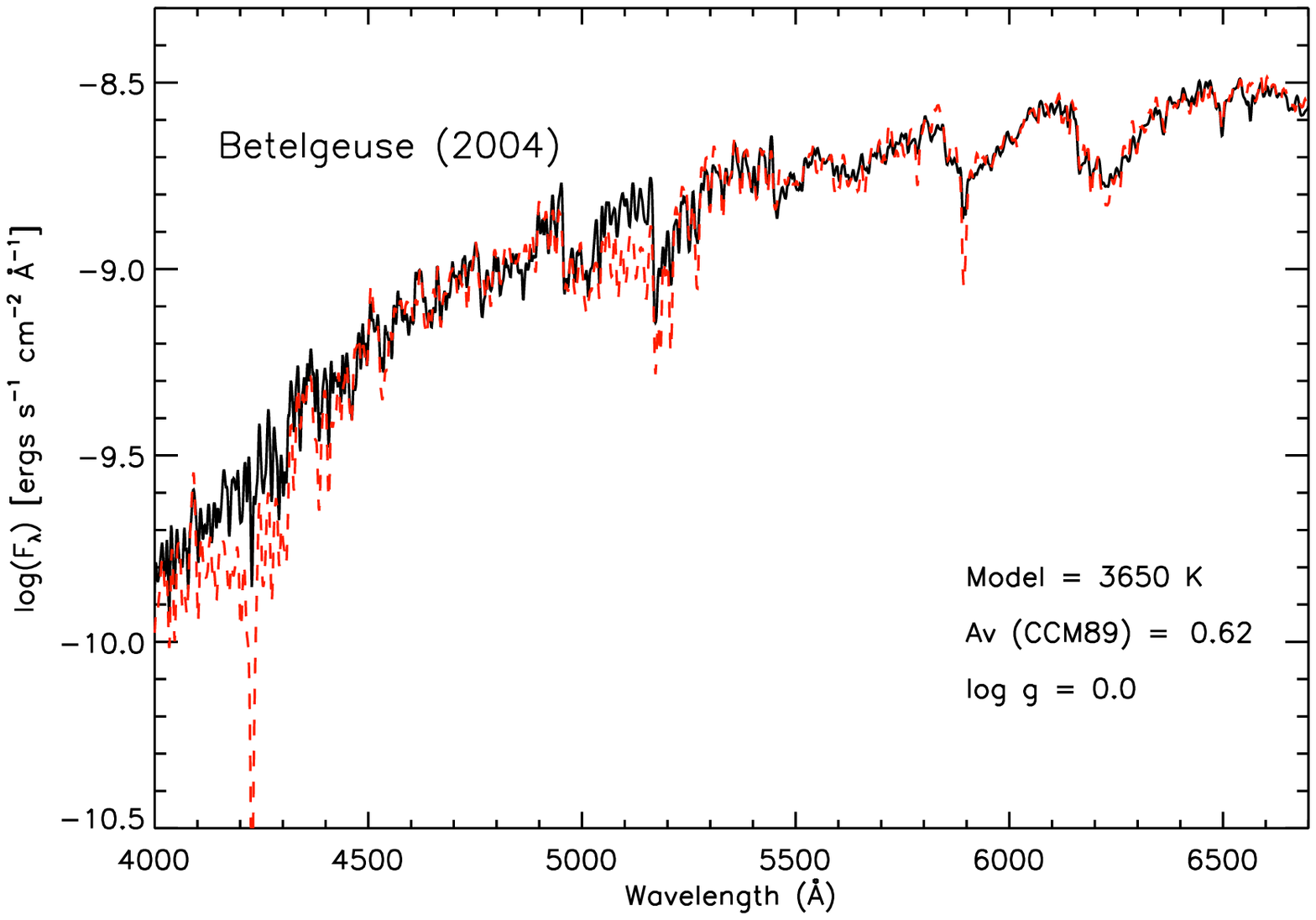}
\plotone{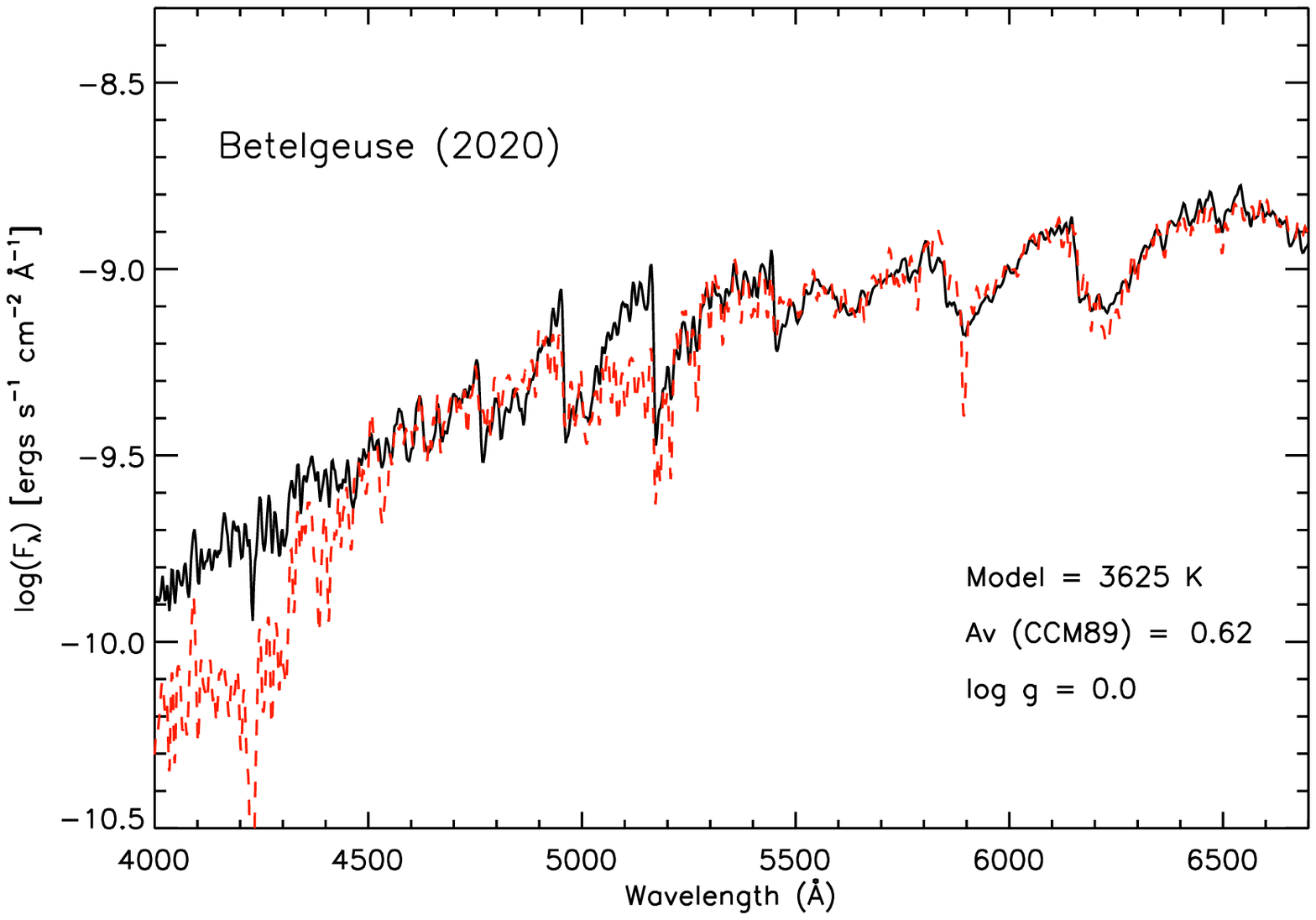}
\caption{Optical spectrophotometry of Betelgeuse from 2004 (left) and 2020 (right), overplotted with the best-fit MARCS stellar atmosphere models (dashed red). The 2004 spectrum is best fit by a 3650 K model, while the 2020 is best fit by a 3625 K model. Note the increased excess flux at $<$4500\AA\ relative to the best-fit MARCS model in the 2020 spectrum.}
\end{figure}

\end{document}